\documentclass[aps,twocolumn,showpacs,preprintnumbers,amsmath,amssymb,superscriptaddress,floatfix,nofootinbib]{revtex4}

\usepackage{graphicx}
\usepackage{epsfig}
\usepackage{epstopdf}
\usepackage{hyperref}
\usepackage{amsmath}
\usepackage{amsfonts}
\usepackage{amssymb}

\begin{document}

\title{$\bar B^0$ decay into $D^0$ and $f_0(500)$, $f_0(980)$, $a_0(980)$, $\rho$ and $\bar B^0_s$ decay into $D^0$ and  $\kappa(800)$, $K^{*0}$}

\author{Wei-Hong Liang}
\email{liangwh@gxnu.edu.cn}
\affiliation{Department of Physics, Guangxi Normal University,
Guilin 541004, China}
\affiliation{Institute of Modern Physics, Chinese Academy of
Sciences, Lanzhou 730000, China}

\author{Ju-Jun Xie} \email{xiejujun@impcas.ac.cn}

\affiliation{Institute of Modern Physics, Chinese Academy of
Sciences, Lanzhou 730000, China} \affiliation{Research Center for
Hadron and CSR Physics, Institute of Modern Physics of CAS and
Lanzhou University, Lanzhou 730000, China} \affiliation{State Key
Laboratory of Theoretical Physics, Institute of Theoretical Physics,
Chinese Academy of Sciences, Beijing 100190, China}

\author{E.~Oset}

\affiliation{Institute of Modern Physics, Chinese Academy of
Sciences, Lanzhou 730000, China} \affiliation{Departamento de
F\'{\i}sica Te\'orica and IFIC, Centro Mixto Universidad de
Valencia-CSIC Institutos de Investigaci\'on de Paterna, Aptdo.
22085, 46071 Valencia, Spain}

\date{\today}

\begin{abstract}
We make predictions for ratios of branching fractions of $\bar B^0$
decays into $D^0$ and the scalar mesons $f_0(500)$, $f_0(980)$,
$a_0(980)$, plus $\bar B^0_s$ decay into $D^0$ and $\kappa(800)$. We
also compare the $\pi^+ \pi^-$ production in the scalar channel with
that observed in the $\rho$ channel and make predictions for the
$\bar B^0_s$ decay into $D^0$ and $K^*(892)$, comparing the strength
of this channel with that of $\kappa(800)$ production. The work is
based on results of the chiral unitary approach where the scalar
resonances are generated from the pseudoscalar-pseudoscalar
interaction. Up to an arbitrary normalization, the mass
distributions and rates for decays into the scalar resonances are
predicted with no free parameters. Comparison with  experimental
data is done when available.
\end{abstract}

\maketitle

\section{Introduction}

The weak decay of $B$ mesons has become an unexpected and most
valuable source of information on hadron structure and in particular
a powerful instrument to investigate the nature of the scalar
mesons, which is a permanent source of debate. The starting point in
this line came with the observation in LHCb \cite{Aaij:2011fx} that
in the  $B^0_s$ decay into $J/\psi$ and $\pi^+ \pi^-$ a pronounced
peak for the $f_0(980)$ was observed, while no signal was seen for
the $f_0(500)$ ($\sigma$). This finding was corroborated by
following experiments by the Belle~\cite{Li:2011pg},
CDF~\cite{Aaltonen:2011nk}, and D0~\cite{Abazov:2011hv}
collaborations. Soon it was also observed that in the $B^0$ decay
into $J/\psi$ and $\pi^+ \pi^-$ \cite{Aaij:2013zpt,Aaij:2014siy}, a
clear signal was seen for $f_0(500)$ production while no signal, or
a very small one, was seen for $f_0(980)$.

The low lying scalar mesons have been the subject of study within
the unitary extension of chiral perturbation theory, the so-called
chiral unitary approach, and a coherent picture emerges where these
states are generated from the interaction of pseudoscalar mesons
provided by the chiral
Lagrangians~\cite{npa,ramonet,kaiser,markushin,juanito,rios}. Some
other approaches use different starting points, like assuming a seed
of $q \bar q$ \cite{van Beveren:1986ea,Tornqvist:1995ay}, or a
tetraquark component~\cite{amir,amirscatt}, but as soon as these
original components are allowed to mix with the unavoidable
meson-meson components, the large strength of this interaction "eats
up" the original seed and the meson-meson cloud becomes the largest
component of the states.

The dynamical picture to generate the scalar mesons from the
pseudoscalar-pseudoscalar interaction has been tested successfully
in a large number of reactions~\cite{review} (see a recent update in
Ref.~\cite{liang}). However, the findings of the $B$ decays have
opened a new line of research on this topic, offering new and useful
information on the structure of these scalar mesons. Indeed, in
Ref.~\cite{liang} it was shown that the features and ratios obtained
from the experiments on $B$ decays could be well reproduced by the
dynamical generation picture of the scalars. It was shown there,
that  although addressing the full complexity of these and related
problems can be rather complicated and require many free
parameters~\cite{robert,bruno,cheng,bruno2,lucio,Colangelo:2010bg,Dedonder:2014xpa},
the evaluation of ratios of decay modes for some of these channels
is rather simple and, in particular, allows one to get an insight on
the structure of the scalar resonances. We shall also mention that
our approach is based on the use of the dominant Cabibbo allowed
decay mechanisms at the quark level. The approach does not contain
subdominant amplitudes which are also considered, for instance, in studies of
$CP$ violation~\cite{ElBennich:2009da,Dedonder:2010fg}, but this is
not our purpose here.

A related but different path is followed in Ref.~\cite{stone},
looking at the scalars from the point of view of $q \bar q$ or
tetraquarks, but no consideration of the final state interaction of
these mesons is done there, while this is at the heart of the
generation of the scalar mesons in the chiral unitary approach.

The work of Ref.~\cite{liang} on $B^0_s$ and $B^0$ decays in
$J/\psi$ and $\pi^+ \pi^-$ has followed suit along the same lines
and in Ref.~\cite{bayarvec} the rates for $B^0_s$ and $B^0$ decays
in $J/\psi$ and a vector meson were investigated and successfully
reproduced, along with predictions for the decays into $J/\psi$ and
$\kappa(800)$. Similarly, in Ref.~\cite{xievec} predictions were
done for the ratios of branching fractions of ${\bar B}^0$ and
${\bar B}^0_s$ decays into $J/\psi$ and the scalar mesons
$f_0(1370)$, $f_0(1710)$, or tensor mesons $f_2(1270)$,
$f'_2(1525)$, $K^*_2(1430)$. Related work, but on weak $D$ decays
into $K^0$ and the $f_0(500)$, $f_0(980)$, and $a_0(980)$ has been
done in Ref.~\cite{daiddec}. One of the interesting things about these
weak decays is that isospin is not conserved and then one can obtain
states of different isospin, like the $f_0(980)$ and $a_0(980)$,
from the same reaction. The prediction for the rates of these two
channels from the same reaction is a new test offered by these weak
decays.

In the present paper we undertake a related problem.  We study the
decay of $\bar B^0$ into $D^0$ and $f_0(500)$, $f_0(980)$, and
$a_0(980)$. At the same time we study the decay of $\bar B^0_s$ into
$D^0$ and $\kappa(800)$. We also relate the rates of production of
vector mesons and compare $\rho$ with $f_0(500)$ production and
$K^{*0}$ with $\kappa(800)$ production. Experimentally there is
information on $\rho$ and $f_0(500)$ production in
Ref.~\cite{Kuzmin:2006mw} for the $\bar B^0$ decay into $D^0$ and
$\pi^+ \pi^-$. There is also information on the ratio of the rates
for $B^0 \to \bar D^0 K^+ K^-$ and $B^0 \to \bar D^0 \pi^+
\pi^-$~\cite{Aaij:2012zka}. We investigate all these rates and
compare them with the experimental information.

\section{Formalism}

\begin{figure}[htbp]
\begin{center}
\includegraphics[scale=0.65]{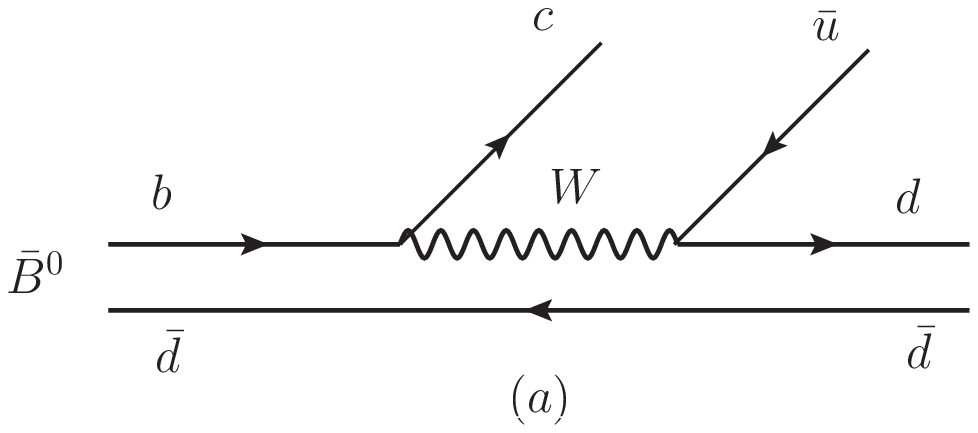}
\includegraphics[scale=0.65]{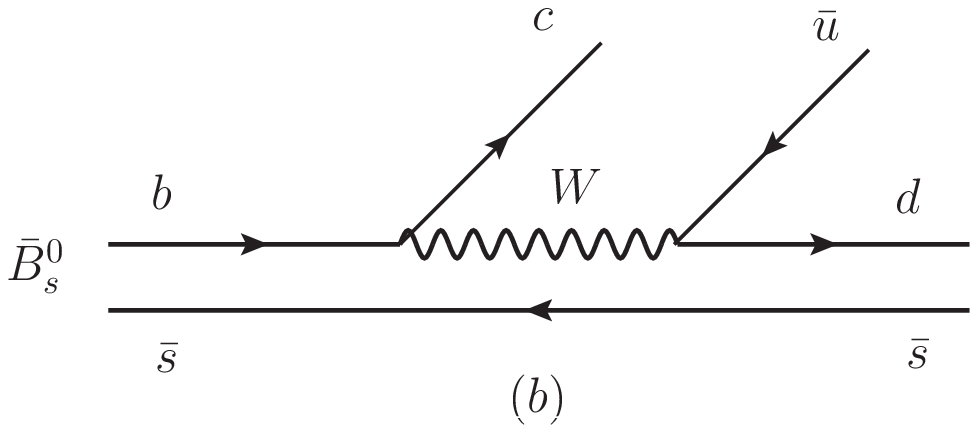}
\caption{Diagrammatic representations of $\bar{B}^0 \to D^0 d
\bar{d}$ decay (a) and $\bar{B}^0_s \to D^0 d \bar{s}$ decay (b).}
\label{Fig:btodqqbar}
\end{center}
\end{figure}

Following Refs.~\cite{liang} and \cite{stone} we show in
Fig.~\ref{Fig:btodqqbar} the dominant diagrams for $\bar{B}^0$
[Fig.~\ref{Fig:btodqqbar} (a)] and $\bar{B}^0_s$
[Fig.~\ref{Fig:btodqqbar} (b)] decays at the quark level. The
mechanism has the $b \to c$ transition, needed for the decay, and
the $u \to d$ vertex that requires the Cabibbo favored $V_{ud}$
Cabibbo-Kobayashi-Maskawa (CKM) matrix element ($V_{ud} =
\cos\theta_c$). Note that these two processes have the same two weak
vertices. Under the assumption that the $\bar{d}$ in
Fig.~\ref{Fig:btodqqbar} (a) and the $\bar{s}$ in
Fig.~\ref{Fig:btodqqbar} (b) act as spectators in these processes,
these amplitudes are identical.

\subsection{$\bar{B}^0$ and $\bar{B}^0_s$ decay into $D^0$ and a vector}

Figure~\ref{Fig:btodqqbar} (a) contains $d\bar{d}$ from where the
$\rho$ and $\omega$ mesons can be formed. Figure~\ref{Fig:btodqqbar}
(b) contains $d\bar{s}$ from where the $K^{*0}$ emerges. At the
quark level, we have
\begin{eqnarray}
&&|\rho^0> = \frac{1}{\sqrt{2}}(u\bar{u} - d\bar{d});~~ |\omega> =
\frac{1}{\sqrt{2}} (u\bar{u} + d\bar{d}); \\
&&|K^{*0}> = d\bar{s}.
\end{eqnarray}

Hence, by taking as reference the amplitude for $\bar{B}^0 \to D^0
K^*$ as $V'_P p_D$, we can write the rest of the amplitudes as
\begin{eqnarray}
&&t_{\bar{B}^0 \to D^0 \rho^0}  =  -\frac{1}{\sqrt{2}} V'_P p_D ,
\label{bzerotodrho} \\
&&t_{\bar{B}^0 \to D^0 \omega}  =  \frac{1}{\sqrt{2}} V'_P p_D ,
\label{bzerotodomega} \\
&&t_{\bar{B}^0 \to D^0 \phi}  = 0 ,
\label{bzerotodphi} \\
&&t_{\bar{B}^0_s \to D^0 K^{*0}}  =   V'_P p_D , \label{bzerotodkstar}
\end{eqnarray}
where $V'_P$ is a common factor to all $\bar B^0 (\bar B^0_s) \to D^0 V_i$ decays, with $V_i$ being a vector meson,
and $p_D$ the momentum of the $D^0$ meson in the rest frame of
the $\bar{B}^0$ (or $\bar{B}^0_s$),
\begin{equation}\label{eq:pD}
 p_D = \frac{\lambda^{1/2}(M^2_{\bar{B}^0_i}, M^2_D, M^2_{V_i})}{2M_{\bar{B}^0_i}}
\end{equation}
where $\lambda$ is the K\"allen function with $\lambda(x, y, z) = (x
- y - z)^2 - 4yz$.

The factor $p_D$ is included to account for a necessary $P$-wave
vertex to allow the transition from $0^- \to 0^- 1^-$. Although
parity is not conserved, angular momentum is, and this requires the angular momentum
$L=1$. Note that the angular momentum needed here is different than
the one in the $\bar{B}^0 \to J/\psi V_i$, where
$L=0$~\cite{bayarvec}. Hence, a mapping from the situation there to
the present case is not possible.

The decay width is given by
\begin{eqnarray}
\Gamma_{\bar B^0_i \to D^0 V_i} = \frac{1}{8\pi M^2_{\bar{B}^0_i}} |t_{\bar{B}^0_i \to
D^0 V_i}|^2 p_D .
\end{eqnarray}

\subsection{$\bar{B}^0$ and $\bar{B}^0_s$ decay into $D^0$ and a pair of pseudoscalar mesons}

\begin{figure}[htbp]
\begin{center}
\includegraphics[scale=0.6]{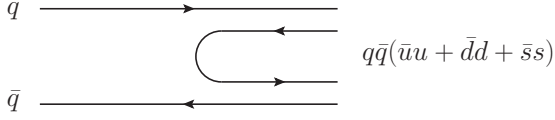}
\caption{Schematic representation of the hadronization of a
$q\bar{q}$ pair.} \label{Fig:qqbarhadronization}
\end{center}
\end{figure}

In order to produce a pair of mesons, the final quark-antiquark pair $d\bar{d}$ or
$d\bar{s}$ in Fig.~\ref{Fig:btodqqbar} has to hadronize into two mesons. The flavor
content, which is all we need in our study, is easily accounted for
in the following way~\cite{alberzou,liang}: we must add a $\bar{q}q$
pair with the quantum numbers of the vacuum, $\bar{u}u + \bar{d}d +
\bar{s}s$, as shown in Fig.~\ref{Fig:qqbarhadronization}.

The content of the meson-meson components in the hadronized
$q\bar{q}$ pair is easily done in the following
way~\cite{alberzou,liang}:
\begin{eqnarray}
M = \left(
           \begin{array}{ccc}
             u\bar u & u \bar d & u\bar s \\
             d\bar u & d\bar d & d\bar s \\
             s\bar u & s\bar d & s\bar s \\
           \end{array}
         \right) = \left(
           \begin{array}{c}
            u   \\
             d  \\
             s   \\
           \end{array}
         \right) \left(
           \begin{array}{ccc}
            \bar{u} & \bar{d} & \bar{s}
           \end{array}   \right),
\end{eqnarray}
where $M$ is the $q\bar{q}$ matrix; then we have the property
\begin{eqnarray}
M \cdot M &=& \left(
           \begin{array}{c}
            u   \\
             d  \\
             s   \\
           \end{array}
         \right) \left(
           \begin{array}{ccc}
            \bar{u} & \bar{d} & \bar{s}
           \end{array}   \right)
           \left(
           \begin{array}{c}
            u   \\
             d  \\
             s   \\
           \end{array}
         \right) \left(
           \begin{array}{ccc}
            \bar{u} & \bar{d} & \bar{s}
           \end{array}   \right) \nonumber \\
           &=& \left(
           \begin{array}{c}
            u   \\
             d  \\
             s   \\
           \end{array}
         \right) \left(
           \begin{array}{ccc}
            \bar{u} & \bar{d} & \bar{s}
           \end{array}   \right) (\bar{u}u + \bar{d}d + \bar{s}s)
           \nonumber \\
           &=& M (\bar{u}u + \bar{d}d + \bar{s}s). \label{MdotM}
\end{eqnarray}

The next step consists of writing the matrix $M$ in terms of mesons
and we have, using the standard $\eta$-$\eta'$
mixing~\cite{bramon,rocapalo},
\begin{widetext}
\begin{equation}\label{eq:phimatrix}
\Phi = \left(
           \begin{array}{ccc}
             \frac{1}{\sqrt{2}}\pi^0 + \frac{1}{\sqrt{3}}\eta + \frac{1}{\sqrt{6}}\eta' & \pi^+ & K^+ \\
             \pi^- & -\frac{1}{\sqrt{2}}\pi^0 + \frac{1}{\sqrt{3}}\eta + \frac{1}{\sqrt{6}}\eta' & K^0 \\
            K^- & \bar{K}^0 & -\frac{1}{\sqrt{3}}\eta + \sqrt{\frac{2}{3}}\eta' \\
           \end{array}
         \right) .
\end{equation}
\end{widetext}
Note that this matrix is different than the standard one used in
chiral theory~\cite{gasser} and used in Ref.~\cite{npa}, from where
we evaluate the meson-meson amplitudes. The difference between the
two matrices is $\frac{1}{\sqrt{3}} {\rm diag} (\eta_1, \eta_1,
\eta_1)$ where $\eta_1$ is the singlet of $SU(3)$, which is
neglected in the matrix used in chiral theory. The reason is that
since the meson-meson interactions are of the type $(\Phi
\partial_{\mu} \Phi - \partial_{\mu} \Phi \Phi)^2$, the singlet
contributions are inoperative there.

Hence, we can write
\begin{eqnarray}
d\bar{d} (\bar{u}u + \bar{d}d + \bar{s}s) \to (\Phi \cdot \Phi)_{22}
&=& \pi^- \pi^+ + \frac{1}{2} \pi^0\pi^0 + \frac{1}{3}\eta\eta
\nonumber \\
&& \!\!\! - \sqrt{\frac{2}{3}}\pi^0 \eta + K^0 \bar{K}^0, \label{phiphi22}\\
s\bar{d} (\bar{u}u + \bar{d}d + \bar{s}s) \to (\Phi \cdot \Phi)_{23}
&=& \! \pi^- K^+ - \! \frac{1}{\sqrt{2}} \pi^0 K^0, \label{phiphi23}
\end{eqnarray}
where we have neglected the terms including $\eta'$ that has too
large mass to be relevant in our study.

\begin{figure*}[htbp]
\begin{center}
\includegraphics[scale=0.5]{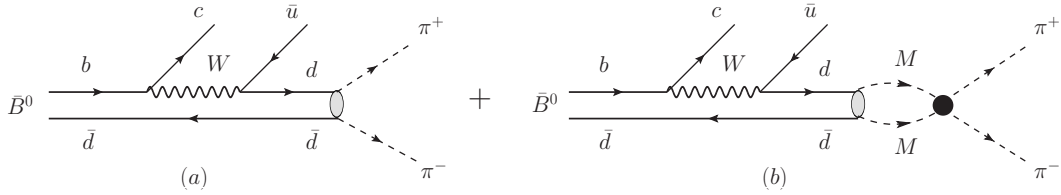}
\caption{Diagrammatic representation of the final state interaction
of the two mesons produced in a primary step. (a) Direct meson-meson production, (b) meson-meson production through rescattering.} \label{Fig:btodpipi}
\end{center}
\end{figure*}

Eqs.~(\ref{phiphi22}) and (\ref{phiphi23}) give us the weight for
pairs of two pseudoscalar mesons. The next step consists of letting
these mesons interact, which they inevitably will do. This is done
in Ref.~\cite{liang} following the mechanism of
Fig.~\ref{Fig:btodpipi}.

The $f_0(500)$ and $f_0(980)$ will be observed in the $\bar{B}^0$
decay into $D^0$ and $\pi^-\pi^+$ final pairs, the $a_0(980)$ in
$\pi^0 \eta$ pairs and the $\kappa(800)$ in the $\bar{B}^0_s$ decay
into $D^0$ and $\pi^- K^+$ pairs. Then we have for the corresponding
production amplitudes
\begin{eqnarray}
&& t(\bar{B}^0 \to D^0 \pi^- \pi^+) = V_P  (1 + G_{\pi^- \pi^+}
t_{\pi^- \pi^+ \to \pi^- \pi^+} \nonumber \\
&& + \frac{1}{2}\frac{1}{2} G_{\pi^0\pi^0} t_{\pi^0\pi^0 \to \pi^-
\pi^+} + \frac{1}{3}\frac{1}{2} G_{\eta \eta} t_{\eta \eta \to \pi^-
\pi^+} \nonumber \\
&& + G_{K^0 \bar{K}^0} t_{K^0 \bar{K}^0 \to \pi^- \pi^+} ),
\label{bzerotodpipi}
\end{eqnarray}
where $V_P$ is a common factor of all these processes, $G_i$ is the
loop function of two meson propagators, and we have included the
factor $\frac{1}{2}$ in the intermediate loops involving a pair of
identical mesons. The elements of the scattering matrix $t_{i \to
j}$ are calculated in Refs.~\cite{liang,daiddec} following the
chiral unitary approach in Refs.~\cite{npa,guome}. Note that the use
of a common $V_P$ factor in Eq.~(\ref{bzerotodpipi}) is related to
the intrinsic $SU(3)$ symmetric structure of the hadronization $\bar
u u + \bar d d + \bar s s$, which implicitly assumes that we add an
$SU(3)$ $\bar q q$ singlet.

Similarly, we can also produce $K^+ K^-$ pairs and we have
\begin{eqnarray}
&& t(\bar{B}^0 \to D^0 K^+ K^-) = V_P  ( G_{\pi^- \pi^+}
t_{\pi^- \pi^+ \to K^+ K^-} \nonumber \\
&& + \frac{1}{2}\frac{1}{2} G_{\pi^0\pi^0} t_{\pi^0\pi^0 \to K^+
K^-} + \frac{1}{3}\frac{1}{2} G_{\eta \eta} t_{\eta \eta \to K^+ K^-} \nonumber \\
&& - \sqrt{\frac{2}{3}}G_{\pi^0 \eta} t_{\pi^0 \eta \to K^+ K^-} +
G_{K^0 \bar{K}^0} t_{K^0 \bar{K}^0 \to K^+ K^-}).~~~~
\label{bzerotodkk}
\end{eqnarray}

In the same way we can write\footnote{It is worth noting that
$\pi^+\pi^-$, $\pi^0 \pi^0$, and $\eta \eta$ are in isospin $I=0$,
while $\pi^0 \eta$ is in $I=1$.}
\begin{eqnarray}
&& t(\bar{B}^0 \to D^0 \pi^0 \eta) = V_P  ( -\sqrt{\frac{2}{3}} -
\sqrt{\frac{2}{3}}G_{\pi^0 \eta} t_{\pi^0 \eta \to \pi^0 \eta}
\nonumber \\
&& + G_{K^0 \bar{K}^0} t_{K^0 \bar{K}^0 \to \pi^0 \eta} ),
\label{bzerotodpieta}
\end{eqnarray}
and taking into account that the amplitude for $\bar{B}^0_s \to c
\bar u + d \bar s$ in Fig.~\ref{Fig:btodqqbar} (b) is the same as
for $\bar{B}^0 \to c \bar u + d \bar d$ of Fig.~\ref{Fig:btodqqbar}
(a), and using Eq.~(\ref{phiphi23}) to account for hadronization, we
obtain
\begin{eqnarray}
&& t(\bar{B}^0_s \to D^0 \pi^- K^+) = V_P  ( 1 + G_{\pi^- K^+}
t_{\pi^- K^+ \to \pi^- K^+} \nonumber \\
&& - \frac{1}{\sqrt{2}} G_{\pi^0 K^0} t_{\pi^0 K^0 \to \pi^- K^+}),
\label{bstodpik}
\end{eqnarray}
where the amplitudes $t_{\pi^- K^+ \to \pi^- K^+}$ and $t_{\pi^0 K^0
\to \pi^- K^+}$ are taken from Ref.~\cite{guome}.

In the process of meson-meson scattering in the $S$-wave, as we shall
study here in order to get the scalar resonances, we have the
transition $0^- \to 0^- 0^+$ for $\bar{B}^0 \to D^0 f_0$, and now we
need $L=0$. Once again the roles of the angular momentum are
reversed with respect to the meson pair production in the $\bar{B}^0
\to J/\psi \pi^+ \pi^-$ decay~\cite{bayarvec}. Hence, we can write
the differential invariant mass width as
\begin{eqnarray}
\frac{d\Gamma}{dM_{\rm inv}} = \frac{1}{(2\pi)^3} \frac{p_D
\tilde{p}_{\pi}} {4M^2_{\bar{B}^0}}\left |t(\bar{B}^0 \to D^0 \pi^-
\pi^+)\right|^2, \label{dgamrdm-pipi}
\end{eqnarray}
where $\tilde{p}_{\pi}$ is the pion momentum for the
$\pi^+$ or $\pi^-$ in the rest frame of the $\pi^- \pi^+$ system
\begin{equation}\label{eq:p-pi}
  \tilde{p}_{\pi} = \frac{\lambda^{1/2}(M^2_{\rm inv}, m^2_{\pi},
m^2_{\pi})}{2M_{\rm inv}},
\end{equation}
where $M_{\rm inv}$ is the invariant mass of the $\pi^+ \pi^-$
system, and also write similar formulas for the other decays.

\section{Numerical results}

In the first place we look for the rates of $\bar{B}^0$ and
$\bar{B}^0_s$ decay into $D^0$ and a vector. By looking at
Eqs.~(\ref{bzerotodrho}), (\ref{bzerotodomega}), and
(\ref{bzerotodkstar}), we have
\begin{eqnarray}
&&\frac{\Gamma_{\bar{B}^0 \to D^0 \rho^0}}{\Gamma_{\bar{B}^0 \to D^0
\omega}}  =  \left[\frac{p_D(\rho^0)}{p_D(\omega)} \right]^3 = 1 , \label{rhotoomega} \\
&&\frac{\Gamma_{\bar{B}^0 \to D^0 \rho^0}}{\Gamma_{\bar{B}^0_s \to D^0
K^{*0}}}  = \left(\frac{M_{\bar{B}^0_s}}{M_{\bar{B}^0}}\right)^2 \frac{1}{2} \left[\frac{p_D(\rho^0)}{p_D(K^{*0})} \right]^3 \simeq \frac{1}{2} , \label{rhotokstar} \\
&&\Gamma_{\bar{B}^0 \to D^0 \phi} = 0 .
\end{eqnarray}

Experimentally there are no data in the PDG~\cite{pdg} for the
branching ratio $Br({\bar{B}^0 \to D^0 \phi})$ and we find the
branching ratios for $B^0 \to \bar{D}^0
\rho^0$~\cite{Kuzmin:2006mw}, $B^0 \to \bar{D}^0 \omega$~\cite{lees,
Blyth}, and $B^0_s \to \bar{D}^0
\bar{K}^{*0}$~\cite{pr1,pr2,Kuzmin:2006mw}, as the following (note
the change $\bar{B}^0 \to B^0$ and $D^0 \to \bar{D}^0$, $\bar{B}^0_s
\to B^0_s$, $K^{*0} \to \bar{K}^{*0}$):
\begin{eqnarray}
Br(B^0 \to \bar{D}^0 \rho^0) &=& (3.2 \pm 0.5) \times 10^{-4}, \label{gamrrhoexp} \\
Br(B^0 \to \bar{D}^0 \omega) &=& (2.53 \pm 0.16) \times 10^{-4}, \label{gamromegaexp} \\
Br(B^0_s \to \bar{D}^0 \bar{K}^{*0}) &=& (3.5 \pm 0.6) \times 10^{-4}.
\label{gamrkstarexp}
\end{eqnarray}

The ratio $\frac{\Gamma_{\bar{B}^0 \to D^0
\rho^0}}{\Gamma_{\bar{B}^0 \to D^0 \omega}}$ is fulfilled, while the
ratio $\frac{\Gamma_{\bar{B}^0 \to D^0 \rho^0}}{\Gamma_{\bar{B}^0_s
\to D^0 K^{*0}}}$ is barely in agreement with data. The branching
ratio of Eq. (\ref{gamrkstarexp}) requires combining ratios obtained
in different experiments. A direct measure from a single experiment
is available in Ref. \cite{exp1}:
\begin{eqnarray}
\frac{\Gamma_{\bar{B}^0_s \to D^0 K^{*0}}}{\Gamma_{\bar{B}^0 \to D^0
 \rho^0}} = 1.48 \pm 0.34 \pm 0.15 \pm 0.12,\label{ratioBtoDexp}
\end{eqnarray}
which is compatible with the factor of $2$ that we get from
Eq.~(\ref{rhotokstar}). However, the result of
Eq.~(\ref{gamrkstarexp}), based on more recent measurements from
Refs.~\cite{pr1} and \cite{pr2}, improve on the result of
Eq.~(\ref{ratioBtoDexp})\cite{TimGershon}, which means that our prediction
for this ratio is a bit bigger than experiment.

\begin{figure*}[htbp]
\begin{center}
\includegraphics[scale=0.35]{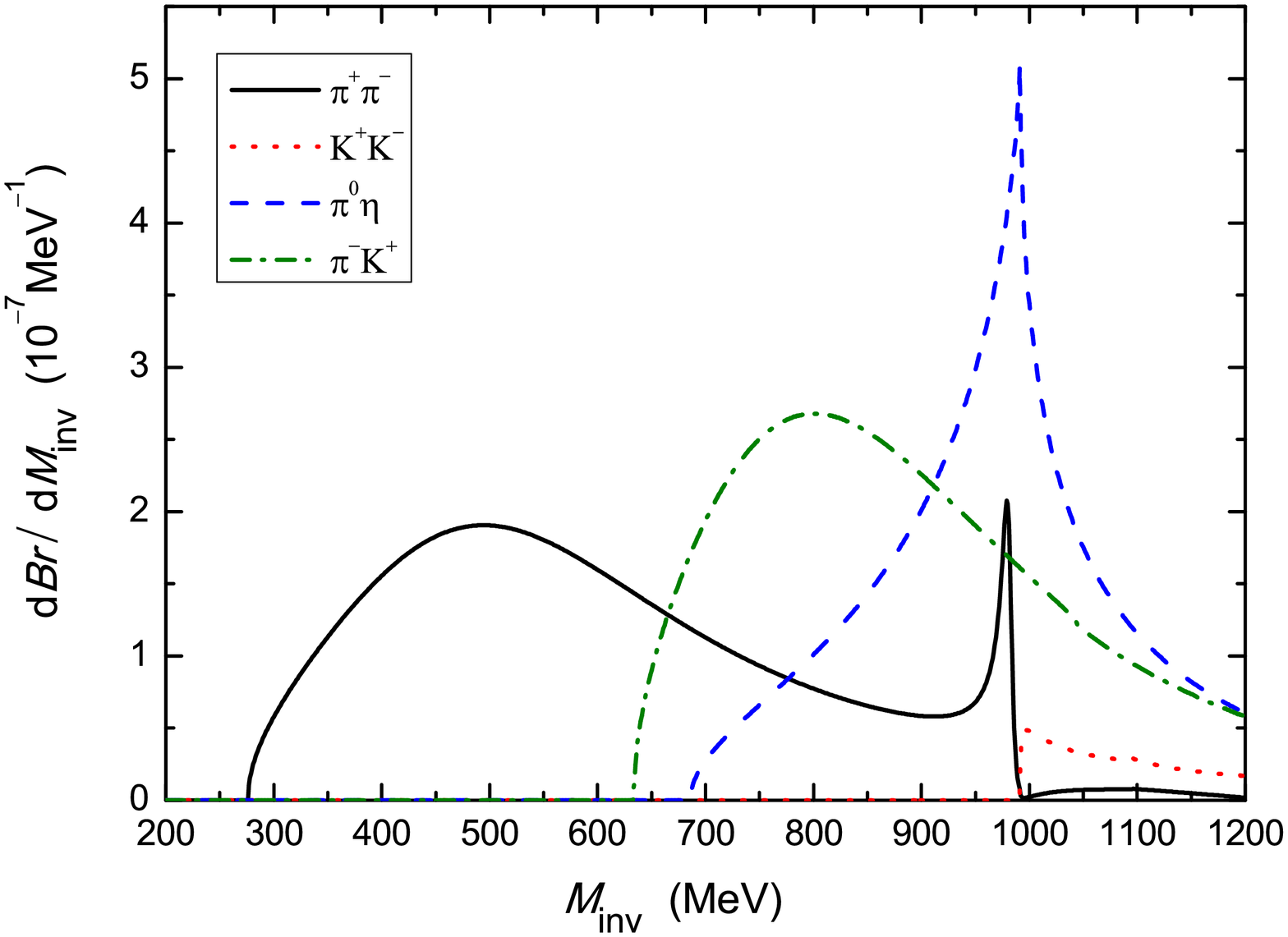}
\caption{ Invariant mass distributions for the
$\pi^+\pi^-$, $K^+ K^-$, and $\pi^0 \eta$, and $\pi^- K$ in
$\bar{B}^0 \to D^0 \pi^+ \pi^-$,~$ D^0 K^+ K^-$, $D^0 \pi^0
\eta$, and $\bar{B}^0_s \to D^0 \pi^- K^+$ decays. The normalization
is such that the integral over the $f_0(500)$ signal gives the experimental branching ratio of Eq. (\ref{sigmabran}).}
\label{Fig:dgamrdminv}
\end{center}
\end{figure*}

We turn now to the production of the scalar resonances. By using
Eqs.~(\ref{bzerotodpipi})-(\ref{bstodpik}),
we obtain the mass
distributions for $\pi^+\pi^-$, $K^+ K^-$, and $\pi^0 \eta$ in
$\bar{B}^0$ decays and $\pi^- K$ in $\bar{B}^0_s$ decay. The
numerical results are shown in Fig.~\ref{Fig:dgamrdminv}.

The normalization for all the processes is the same. The scale is obtained demanding that the integrated $f_0(500)$ distribution has the normalization of the experimental branching ratio of
Eq. (\ref{sigmabran}). From
Fig.~\ref{Fig:dgamrdminv}, in the $\pi^+ \pi^-$ invariant mass
distribution for $\bar B^0 \to D^0 \pi^+ \pi^-$ decay, we observe an
appreciable strength for $f_0(500)$ excitation and a less strong,
but clearly visible excitation for the $f_0(980)$. In the $\pi^0 \eta$
invariant mass distribution, the $a_0(980)$ is also excited with a
strength bigger than that of the $f_0(980)$. Finally, in the $\pi^-
K^+$ invariant mass distribution, the $\kappa(800)$ is also excited
with a strength comparable to that of the $f_0(500)$. We also plot
the mass distribution for $K^+ K^-$ production. It begins at
threshold and gets strength from the two underlying $f_0(980)$ and
$a_0(980)$ resonances, hence we can see an accumulated strength
close to threshold that makes the distribution clearly different
from phase space.

There is some experimental information to test some of the
predictions of our results. Indeed in Ref.~\cite{Kuzmin:2006mw} (see
Table II of that paper) one can find the rates of production for
$f_0(500)$ [it is called $f_0(600)$ there] and $f_0(980)$.
Concretely,
\begin{eqnarray}
&&Br[\bar B^0 \to D^0 f_0(500)] \cdot Br[f_0(500) \to \pi^+ \pi^-] \nonumber \\
&=& (0.68 \pm 0.08) \times 10^{-4}, \label{sigmabran} \\
&&Br[\bar B^0 \to D^0 f_0(980)] \cdot Br[f_0(980) \to \pi^+ \pi^-]  \nonumber \\
&=& (0.08 \pm 0.04) \times 10^{-4},
\end{eqnarray}
where the errors are only statistical. This gives
\begin{eqnarray}
&&\left. \frac{Br[\bar B^0 \to D^0 f_0(980)] \cdot Br[f_0(980)\to \pi^+ \pi^-]}{Br[\bar B^0 \to D^0 f_0(500)] \cdot Br[f_0(500)\to \pi^+ \pi^-]} \right|_{\rm Exp.} \nonumber \\
&=& 0.12 \pm 0.06. \label{eq:ratioBtof0Exp}
\end{eqnarray}

From Fig.~\ref{Fig:dgamrdminv} it is easy to estimate our
theoretical results for this ratio by integrating over the peaks of
the $f_0(500)$ and $f_0(980)$. To separate the $f_0(500)$ and $f_0(980)$ contributions, a smooth extrapolation of the curve of Fig. \ref{Fig:dgamrdminv} is made from 900 to 1000 MeV, as done in Ref.~\cite{daiddec}. We find
\begin{equation}\label{eq:ratioBtof0Theo}
\left. \frac{Br[\bar B^0 \to D^0 f_0(980)] \cdot Br[f_0(980) \to \pi^+ \pi^-]}{Br[\bar B^0 \to D^0 f_0(500)] \cdot Br[f_0(500)\to \pi^+ \pi^-]} \right|_{\rm Theo.}=0.08,
\end{equation}
with an estimated error of about 10\%. As we can see, the agreement
of the theoretical results with experiment is good within errors.

We have selected $\bar{B}^0$ decay into $D^0$  and $\pi^+ \pi^-$ or
$\pi^0 \eta$ and $\bar{B}^0_s$ into $D^0$  and $\pi^- K^+$ which are
Cabibbo favored. In this case one does not find competitive
mechanisms corresponding to different topologies of the diagrams
\cite{Chau1983}. Similarly as done in Ref.~\cite{liang}, one could
also consider $\bar{B}^0_s$ into $D^0$ and $\pi^+ \pi^-$. In this
case we can have this reaction using the mechanism of Fig.
\ref{Fig:btodqqbar}(b), replacing the final $d$ quark with an $s$
quark. Upon hadronization the $s \bar s$ pair will give $K \bar K$, 
which upon rescattering can produce $\pi^+ \pi^-$.  
The $udW$ transition is replaced by the $usW$ transition and hence the $\cos
{\theta_c}$ into $\sin {\theta_c}$. The evaluation of this diagram
is straightforward, but there is a competing diagram of the type of
external emission [see Fig. 5 (a) of Ref.~\cite{daiddec}] where the
$W$ directly converts into $s \bar u (K^-)$ and the final quark is a
$c$ quark. Upon hadronization of the $c \bar s$ pair we can get
$D^0$ and $K^+$.  In both mechanisms we have $K \bar K D^0$ in the
final state, which through rescattering will give $D^0 \pi^+ \pi^-$,
and the two mechanisms interfere. We thus cannot be as predictive as
in the other cases where there is only one dominant mechanism and
unknown dynamical factors cancel in ratios. However, we can already
say that these two mechanisms are both Cabibbo suppressed, so the
ratio of $f_0(980)$ production in this case would be suppressed with
respect to the $B^0$ case by $(\sin \theta_c / \cos \theta_c)^2$
with respect to the $B^0$ case. This is in contrast to the $B^0$ and
$B^0_s$ decays into $J/\psi$ and $f_0(980)$, where the second decay
was favored with respect to the first
one~\cite{Aaij:2011fx,Li:2011pg,Aaltonen:2011nk,Abazov:2011hv,Aaij:2013zpt,Aaij:2014siy,liang}.
On the other hand, we see also here that the $\pi^+ \pi^-$ in the
$\bar{B}^0_s$ decay into $D^0$  and $\pi^+ \pi^-$ proceeds via
rescattering of the primary produced $K \bar K $ pair.  This is
similar to the case of $B^0_s$ decay into $J/\psi$ and $ \pi^+
\pi^-$ in Ref.~\cite{liang}, and thus we can also predict that in the
$\bar{B}^0_s \to D^0 \pi^+ \pi^-$ the $f_0(980)$, although Cabibbo
suppressed, could be seen and there would be practically no trace of
the $f_0(500)$ excitation.

It is most instructive to show the $\pi^+ \pi^-$ production combining the $S$-wave and
$P$-wave production. In order to do that, we evaluate $V_P$ of Eq. (\ref{bzerotodpipi}) and $V'_P$ of Eq. (\ref{bzerotodrho}), normalized to obtain the branching fractions
given in Eqs.~(\ref{sigmabran}) and (\ref{gamrrhoexp}), rather than
widths. We shall call the parameters  $\tilde{V}_P$ and $\tilde{V}'_P$, suited to this normalization.

We obtain $\tilde{V}_P = (8.8 \pm 0.5) \times 10^{-2} ~{\rm MeV}^{-1/2}$ and
 $\tilde{V}'_P = (6.8 \pm 0.5) \times 10^{-3} ~{\rm MeV}^{-1/2}$.

To obtain the $\pi^+ \pi^-$  mass distribution for the $\rho$, we need to convert
the total rate for vector production into a mass distribution. This
we do by following the steps of Ref.~\cite{bayarvec}, and then we write
\begin{eqnarray}
&&\frac{d\Gamma_{\bar{B}^0 \to D^0 \rho^0 \to D^0 \pi^+
\pi^-}}{dM_{\rm inv}} =  - \frac{2m_{\rho}}{\pi} \times \nonumber \\
&&  {\rm Im}\left[ \frac{1}{M^2_{\rm inv} -m^2_{\rho} + i m_{\rho}
\Gamma_{\rho}(M_{\rm inv})}\right]  \tilde{\Gamma}_{\bar{B}^0 \to D^0
\rho^0}, \label{dgamrdm-rhotopipi}
\end{eqnarray}
where
\begin{eqnarray}
&&\Gamma_{\rho}(M_{\rm inv}) = \Gamma_{\rho} \left ( \frac{p^{\rm
off}_{\pi}}{p^{\rm on}_{\pi}} \right )^3 \frac{m^2_{\rho}}{M^2_{\rm
inv}}, \\
&&p^{\rm off} = \frac{\lambda^{1/2} (M^2_{\rm inv}, m^2_{\pi},
m^2_{\pi})}{2M_{\rm inv}} \theta(M_{\rm inv} - 2m_{\pi}), \\
&& p^{\rm on} = \frac{\lambda^{1/2} (m^2_{\rho}, m^2_{\pi},
m^2_{\pi})}{2m_{\rho}} , \\
&& \tilde{\Gamma}_{\bar{B}^0 \to D^0 \rho^0} (M_{\rm inv})  =
\Gamma_{\bar{B}^0 \to D^0 \rho^0} \left ( \frac{p^{\rm
off}_D}{p^{\rm on}_D} \right )^3
\end{eqnarray}
with $p^{\rm off}_D$ the $D^0$ momentum for $\pi^+ \pi^-$ invariant
mass $M_{\rm inv}$ and $p^{\rm on}_D$ for $M_{\rm inv} = m_{\rho}$.
In order to get the $\pi^- K^+$ mass distribution for $\bar{B}^0_s
\to D^0 \pi^- K^-$,  we apply the
same procedure, changing the mass and width of the vector, and $\pi
K$ instead of $\pi \pi$ in the formula of the width.

\begin{figure}[htbp]
\begin{center}
\includegraphics[scale=0.35]{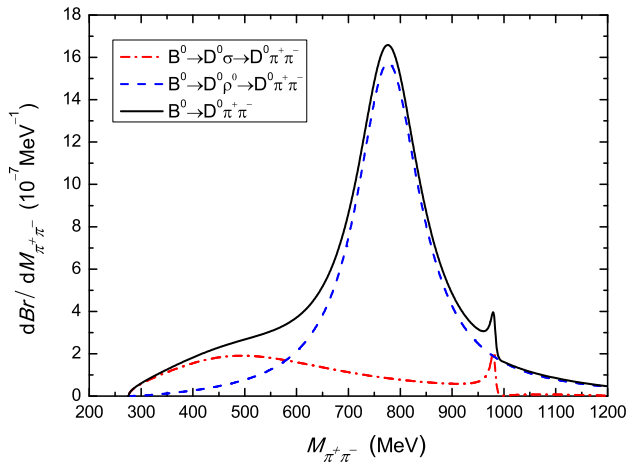}
\caption{Invariant mass distribution for $\pi^+\pi^-$
in $\bar{B}^0 \to D^0 \pi^+ \pi^-$ decay. The normalization is the same as in Fig. \ref{Fig:dgamrdminv}.} \label{Fig:dgamrdminvpipi}
\end{center}
\end{figure}

The formulas are easily generalized for the other decays.

Now we show the results for the $\pi^+ \pi^-$ production in $\bar{B}^0 \to D^0 \pi^+ \pi^-$  in Fig.~\ref{Fig:dgamrdminvpipi}. We see a large contribution from
the $f_0(500)$ and a larger contribution from the $\rho^0 \to \pi^+
\pi^-$ production.  We can see that the $f_0(500)$ is clearly visible in
the distribution of $\pi^+ \pi^-$ invariant mass in the region of
$400 \sim 600$ MeV.

The results of Fig. \ref{Fig:dgamrdminvpipi} cannot be directly compared with the experimental ones of Fig. 5 of Ref. \cite{Kuzmin:2006mw} because in the experiment a cut for events with $\pi \pi$ helicity angles with $\cos(\theta_h)>0$ has been implemented. We cannot evaluate the helicity angles because our procedure to get the $\rho$ signal does not explicitly use the pions. Nevertheless, and with this caveat, the shape of the $\pi \pi$ mass distribution obtained here is remarkably similar to the one of that figure.

\begin{figure}[htbp]
\begin{center}
\includegraphics[scale=0.35]{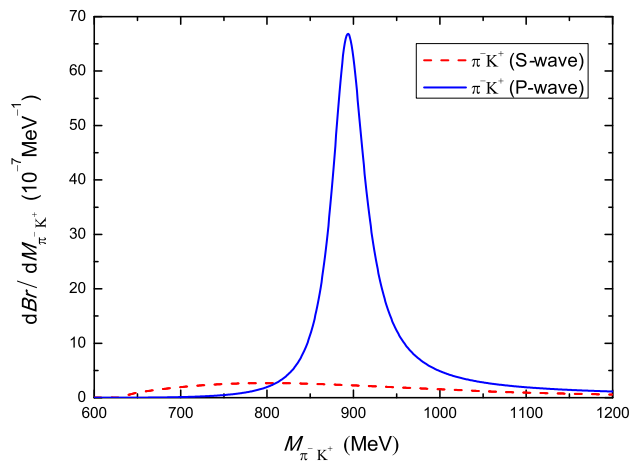}
\caption{Invariant mass distribution for $\pi^- K^+$
in $\bar{B}^0_s \to D^0 \pi^- K^+$ decay. The normalization
is the same as in Fig. \ref{Fig:dgamrdminv}.}
\label{Fig:dgamrdminvpika}
\end{center}
\end{figure}

The $V_P$ and $V'_P$ obtained by fitting the branching ratios of $f_0(500)$ and $\rho$ production
 can be used to obtain the strength of
$K^{*0}$ production versus $\kappa(800)$ production in the $\bar{B}^0_s \to D^0 \pi^- K^+$ decay.
For this we use Eqs.~(\ref{bzerotodrho})-(\ref{bzerotodkstar})
and recall that the
rate for $K^{*0} \to \pi^- K^+$ is $\frac{2}{3}$ of the total
$K^{*0}$ production. The results for $K^{*0} \to \pi^- K^+$ and
$\kappa(800) \to \pi^- K^+$ production are shown in
Fig.~\ref{Fig:dgamrdminvpika}, where we see a clear peak for
$K^{*0}$ production, with strength bigger than that for $\rho^0$ in
Fig.~\ref{Fig:dgamrdminvpipi}, due in part to the factor-of-2 bigger strength in Eq.~(\ref{rhotokstar}) and the smaller $K^{*0}$
width. The $\kappa(800)$ is clearly visible in the lower part of the
spectrum where the $K^{*0}$ has no strength.

Finally, although with more uncertainty, we can also estimate the
ratio
\begin{eqnarray}
\frac{\Gamma(B^0 \to \bar{D}^0 K^+ K^-)}{\Gamma(B^0 \to \bar{D}^0
\pi^+ \pi^-)} = 0.056 \pm 0.011 \pm 0.007 \label{kktopipi}
\end{eqnarray}
of Ref.~\cite{Aaij:2012zka}. This requires an extrapolation of our
results to higher invariant masses where our results would not be
accurate, but, assuming that most of the strength for both reactions
comes from the region close to the $K^+ K^-$ threshold and from the
$\rho^0$ peak, respectively, we obtain a ratio of the order of $0.03
\sim 0.06$, which agrees qualitatively with the ratio of
Eq.~(\ref{kktopipi}).

\section{Conclusions}

In this paper we have addressed the study of the $\bar B^0$ decay
into $D^0$ and $\rho$ or $f_0(500)$, $f_0(980)$, $a_0(980)$, and
$\bar B^0_s$ decay into $D^0$ and $K^*(892)$ or $\kappa(800)$. The
model used is simple to interpret and allows us to get relative
strengths of the different reactions. The Cabibbo favored dominant
mechanism at the quark level is identified and then the rates for
production of vector mesons are trivially obtained assuming a $q
\bar q$ nature for the light vector mesons. The relative rates obtained
are in good agreement with experimental data. This in itself is
already a good finding, supporting the $q\bar q$ structure for the light vector
mesons, which has been advocated from the large $N_c$ behavior of
the amplitudes~\cite{pelaeznc} and from the compositeness sum
rule~\cite{acetirho,acetiks}. As to the production of the scalar
mesons we could predict the invariant mass distributions, up to a
common  global factor, for the $\bar B^0$ decay into $D^0 f_0(500)
[f_0(500) \to \pi^+ \pi^-]$, $D^0 f_0(980) [f_0(980) \to \pi^+
\pi^-]$, $D^0 a_0(980) [a_0(980) \to \pi^0 \eta]$, and $\bar B^0_s$
decay into $D^0 \kappa(800) [\kappa(800) \to \pi^- K^+]$.  Hence the relative weights of the distributions are predicted
with no free parameters under the assumption that
these resonances are generated dynamically from the meson-meson
interactions, and constitute interesting predictions for future
experiments, which are most likely to be performed at LHCb or other
facilities.\footnote{While in the refereeing process two
experimental papers were submitted to the
arXiv~\cite{expnew1,expnew2}. In Ref.~\cite{expnew1} the $B^0 \to
\bar{D}^0 \pi^+ \pi^-$ decay was analyzed and the $\bar{D}^0
f_0(500)$ and $\bar{D}^0 f_0(980)$ modes were observed. It is easy
to see that the ratio of these two branching ratios agree with our
results within errors, and so do the ratios of each of them to
$\rho$ production. In Ref.~\cite{expnew2} the $B^0_s \to \bar{D}^0
f_0(980)$ signal, which we discussed is Cabibbo suppressed, was
found to be very small, and finally an upper limit was provided. On the
other hand, the $B^0_s \to \bar{D}^0 f_0(500)$ mode, which we
predict should not be seen, was not observed.}

We would like to abound in this latter comment. The work done here
follows a different pattern than the one done in many works in
related $B$ decays on
mesons~\cite{robert,bruno,cheng,bruno2,lucio,Colangelo:2010bg,ElBennich:2009da,Dedonder:2010fg}.
These papers address explicitly the dynamics of the weak decays, and
subsequent strong interaction involved in the quark matrix elements,
which are usually evaluated under the factorization approximation.
What makes our work different from other related works, such as 
Ref.~\cite{Dedonder:2014xpa} and similar ones, is that we explicitly
allow the formation of all meson-meson coupled channels in the weak
processes and then allow these meson pairs interact. The resonances
investigated are automatically produced since in our approach it is
precisely the interaction that creates these resonances (dynamical
generation). In Ref.~\cite{Dedonder:2014xpa} and related works, some
channels, as $K \bar K$ in the study of $\pi \pi$ production, are
automatically incorporated by means of form factors at the price of
introducing unknown multiplicative factors to be fitted to the data.
These form factors contain the dynamics of the interaction of the
mesons. Then, different factors appear when using the $K \pi$ or
$\pi \pi$ scalar form factors, but in our approach we could relate
some processes, like the $\bar{B}^0 \to D^0 \pi \pi$ and
$\bar{B}^0_s \to D^0 K \pi$, using a unique unknown factor, $V_P$.
These different approaches are complementary. As mentioned in the
Introduction, our approach, relying on one dominant mechanism,
allows to obtain many ratios with no free parameters, but it cannot be
used to study processes like $CP$ violation which require
at least two weak amplitudes, for which approaches like those of
Refs.~\cite{ElBennich:2009da,Dedonder:2010fg} are demanded. Our
approach is particularly suited to study scalar meson production in
cases where we are confident that these states are dynamically
generated, and the success of our predictions gives further strength
to this hypothesis.

On the other hand,  with the information obtained for $f_0(500)$ and $\rho$ production
and using the experimental rates for these processes, we could make
predictions for the
strength of $K^{*0}$ production in $\bar B^0_s$ decay into $D^0$ and
$K^{*0}$ and compare it with the $\kappa(800)$ contribution. These
are again interesting predictions for future experiments, relative
to the production of the $\rho^0$ in the $\bar B^0$ decay into $D^0$
and $\rho$.

The large amount of information predicted in decays which are
Cabibbo favored, and the relevance that this information has on the
structure of the scalar mesons, should be a clear motivation for the
implementation of these experiments in the near future.

\section*{Acknowledgments}

We would like to thank Dr. Tim Gershon for useful information to
interpret the data and Dr. J. Oller for useful discussions. One of
us, E. O., wishes to acknowledge support from the Chinese Academy of
Science (CAS) in the Program of Visiting Professorship for Senior
International Scientists (Grant No. 2013T2J0012). This work is
partly supported by the Spanish Ministerio de Economia y
Competitividad and European FEDER funds under the contract number
FIS2011-28853-C02-01 and FIS2011-28853-C02-02, and the Generalitat
Valenciana in the program Prometeo II-2014/068. We acknowledge the
support of the European Community-Research Infrastructure
Integrating Activity Study of Strongly Interacting Matter (acronym
HadronPhysics3, Grant Agreement n. 283286) under the Seventh
Framework Programme of EU. This work is also partly supported by the
National Natural Science Foundation of China under Grant Nos.
11165005, 11105126 and 11475227.

\bibliographystyle{plain}

\end{document}